\documentclass[]{llncs}
\PassOptionsToPackage{hyphens}{url}
\usepackage[hidelinks]{hyperref}

\usepackage{packages}
\usepackage[numbers,sort&compress]{natbib}
\begin{document}

\title{QUIC Hunter: Finding QUIC Deployments and Identifying Server Libraries Across the Internet}
\titlerunning{QUIC Library Hunter}
\author{Johannes Zirngibl \and Florian Gebauer \and Patrick Sattler \and Markus Sosnowski \and Georg Carle}
\authorrunning{Zirngibl et al.}
\institute{Technical University of Munich, Germany\\
\email{\{zirngibl, gebauer, sattler, sosnowski, carle\}@net.in.tum.de}}
\maketitle
\begin{abstract}
    The diversity of \quic{} implementations poses challenges for Internet measurements and the analysis of the \quic{} ecosystem.
    While all implementations follow the same specification and there is general interoperability, differences in performance, functionality, but also security (\eg{} due to bugs) can be expected.
    Therefore, knowledge about the implementation of an endpoint on the Internet can help researchers, operators, and users to better analyze connections, performance, and security.
    In this work, we improved the detection rate of \quic{} scans to find more deployments and provide an approach to effectively identify \quic{} server libraries based on \texttt{CONNECTION\_CLOSE} frames and transport parameter orders.
    We performed Internet-wide scans and identified at least one deployment for 18 \quic{} libraries.
    In total, we can identify the libraries with \sm{8.0} IPv4 and \sm{2.5} IPv6 addresses.
    We provide a comprehensive view of the landscape of competing \quic{} libraries.
\end{abstract}

\section{Introduction}%
\label{sec:introduction}

Based on UDP, the new \quic{} protocol~\cite{rfc9000} can be implemented in user space and has thus seen wide attention from various implementors~\cite{quicImplementations2021}.
Several operators (\eg{} Cloudflare~\cite{impl_quiche-cloudflare}), as well as open source teams, started to implement \quic{} libraries during the standardization process.
They regularly update their implementations to follow new developments and improve their library.

Even though all libraries follow the same standard, implementation differences are to be expected.
Related work has shown that these differences affect functionality~\cite{seemann2020automating, marx2020implementationdiversity} and performance~\cite{tyunyayev2022picoquichighspeed,yu2021quicproductionperformance,Jaeger2023QUIC}.
They can drastically influence scans, \eg{} during deployment detection as shown in this work, and future research.
It is important to differentiate between effects due to the network, the protocol specification, and implementation specifics.
The assessment of performance and bottlenecks of \quic{} connections on the Internet~\cite{bauer2023performancebenefits} or new \quic{} features~\cite{sander2023ecn,kunze2023spin} could be improved considering the involved libraries.
Therefore, means to properly scan and identify \quic{} libraries and an overview of their deployment on the Internet are essential.
Our key contributions in this work are:

\first{} We analyze current \quic{} scanning approaches and propose a new \zmap{} approach to identify  more deployments.
We evaluate the current state of \quic{} deployments and analyze the importance of \ac{sni} values. 

\second{} We propose an approach to identify \quic{} libraries based on error messages and transport parameters.
We conducted Internet-wide scans and found at least one deployment for 18 libraries respectively, and 12 different libraries in one \ac{as}.

\third{} We developed a test environment for scanners and our library identification approach.
We published all tools to evaluate future changes of QUIC libraries and scan configurations and to update our approach to identify libraries in case of changes: \url{https://github.com/quic-hunter/libraries}

\section{Background}%
\label{sec:background}

\quic{} combines functionality from different layers, \eg{} transport functionality, but also security through \ac{tls}~\cite{rfc9000}.
It is used on top of \ac{udp} and implemented in user space.
This has led to many \quic{} libraries.
The QUIC working group lists more than 24 different libraries~\cite{quicImplementations2021}, of which 16 are tested by the \quicInteropRunner~\cite{seemann2020automating}.

The \quic{} handshake combines the transport handshake with a \ac{tls} handshake and exchanges various information.
\quic{} restricts TLS to version 1.3 but adds a new \texttt{quic\_transport\_parameters} extension to \ac{tls}~\cite{rfc9001}.
This extension allows the peers to exchange transport parameters during the \quic{} handshake.
These include values like maximum timeouts or limits on the amount of data that may be sent in this connection~\cite{rfc9000}.

Furthermore, relying on \ac{tls} allows \quic{} clients to send a domain as \ac{sni} value during the handshake, indicating the requested service.
\ac{sni} is part of the TLS 1.3 standard and allows a server with multiple domains to select the correct certificate for authentication.
Similarly, \ac{tls} offers an extension called \ac{alpn}, which allows the negotiation of an application layer protocol during the handshake (\eg{} \texttt{h3} to request \ac{http} version 3).
The client sends a list of supported and requested values, while the server selects an offered value or terminates the connection early with an error.

In case of an error, \quic{} peers should send a \texttt{CONNECTION\_CLOSE} frame and terminate the connection.
This frame may contain an error code to explicitly indicate the associated error if there was one.
Interestingly, the frame may include an arbitrary reason phrase provided by the peer and encoded as a string~\cite{rfc9000}.

\textit{We rely on information exchanged during the \texttt{CONNECTION\_CLOSE} frame and \quic{} handshake to identify implementations (see \Cref{sec:identification}).}

\section{Related Work}%
\label{sec:related}

Differences in implementations and their behavior during communications have been used by \citeauthor{sosnowski2022tlsfingerprinting}~\cite{sosnowski2022tlsfingerprinting,Sosnowski2023dissectls} and \citeauthor{jarm}~\cite{jarm} to differentiate TLS/TCP deployments or to analyze operating systems and \ac{tcp} behavior~\cite{nmap2009,Shamsi2016hershelOSfingerprinting}.
In contrast, we focus on \quic{} libraries and specific features independent of the used \ac{tls} library and deployment configurations.

\quic{} deployments and their behavior have been analyzed at different stages of the standardization phase~\cite{ruethQUIC2018, zirngibl2021over9000} and since the release of RFC 9000~\cite{nawrockiTlsCertificates,nawrockiQuicSand}.
\citeauthor{zirngibl2021over9000}~\cite{zirngibl2021over9000} developed and published a \zmap{} module to detect \quic{} implementations and the \qscanner{}.
The \qscanner{} is a stateful \quic{} scanner based on \quicGo{}~\cite{impl_quic-go}.
It allows conducting complete RFC 9000~\cite{rfc9000} conform \quic{} handshakes and HTTP/3 requests.
It extracts \quic{} and \ac{tls} properties, X.509 certificates, and HTTP headers.
According to the \quicInteropRunner~\cite{seemann2020automating}, \quicGo{} can conduct handshakes with all other implementations.
\citeauthor{zirngibl2021over9000} identify widespread deployment mainly driven by large \acp{cdn} and show differences in configured parameters and used versions.
However, they did not analyze which \quic{} libraries are actually used by found deployments.
We base our work on their tools, show an increase in deployments compared to their findings, and provide a methodology to identify the library of found deployments.

\citeauthor{nawrockiTlsCertificates}~\cite{nawrockiTlsCertificates} evaluated the behavior of different deployments with a focus on certificates and their conformance to amplification limits during the handshake.
They show differences between \quic{} deployments.
However, they did not focus on specific library differences and their identification but on the interplay between \quic{} and certificate (chain) sizes.
\citeauthor{muecke2022waitingforquic}~\cite{muecke2022waitingforquic} evaluate \quic{} deployments but focus on specific hyper giants and do not focus on an identification of libraries in general.
Furthermore, \citeauthor{marx2020implementationdiversity}~\cite{marx2020implementationdiversity} have analyzed libraries and showed the impact of different draft interpretations on the implementations, \eg{} due to congestion control mechanisms.
\citeauthor{quicForge}~\cite{quicForge} have evaluated \quic{} libraries focusing on security aspects in local tests.
Other works on \quic{} mainly focused on interoperability~\cite{seemann2020automating,PirauxQuicEvolution2018} and performance~\cite{yang2020quicnicoffloading,shreedhar2022quicperformancewebandstorage,yu2021quicproductionperformance,volodina2021ackscaling,nepomuceno2018QUICandTCP,sander2022resourceprioritization,tyunyayev2022picoquichighspeed,wolsing2019performance, Jaeger2023QUIC}.

While most of these works showed differences in \quic{} libraries and their impact on specific features, none analyzes the impact of these differences on scans and provides means to identify libraries on the Internet. To the best of our knowledge, there is no published test environment for \quic{} scanners and no methodology to identify deployed \quic{} libraries. 
\section{Test Environment}%
\label{sec:environment}

\quic{} scanners are typically built on a single library and assume the correct implementation of the \quic{} protocol as defined by the RFCs.
However, due to the number of available implementations, some implement standards incorrectly, interpret parts of it differently or miss some functionality.
While the \quicInteropRunner{}~\cite{seemann2020automating} covers a variety of test cases between libraries, no test environment was available for scanners.
Therefore, we developed a local test environment to evaluate \quic{} libraries and their behavior to various requests.
It allows for ethical evaluations without interfering with the network and existing deployments in case of unexpected behavior in edge-case scenarios.

\begin{figure}
    \centering
    \footnotesize
    \begin{tikzpicture}[>=latex]%
        \node[font=\fontsize{20}{20},label=above:{\tiny 193.167.100.100},label=left:{Server 1}] (server1) at (0,2.5) {\faServer};
        \node[font=\fontsize{20}{20},label=above:{\tiny 193.167.100.101},label=left:{Server 2}] (server2) at (0,1.5) {\faServer};
        \node[font=\fontsize{20}{20},label=above:{\tiny 193.167.100.1??},label=left:{Server N}] (server3) at (0,0.5) {\faServer};
        \node[font=\fontsize{20}{20}] (router) at (2,1.5) {\faCloud};
        \node[font=\fontsize{20}{20},label=above:{\tiny 193.167.100.90},label=right:{Scanner}] (scanner) at (4,1.5) {\faServer};

        \node[font=\fontsize{20}{20},label= below:{Docker Hub}, left = 100 pt of server1] (hub) {\faServer}; 
        \draw[-latex, dashed,] (hub) -- node[above] {{\lsquic}} ++(2.65,0) ;
        \draw[-latex, dashed,] (hub) -- node[above,sloped] {{\mvfst}} ++(2.65,-1) ;

        \node[font=\fontsize{20}{20},label=above:{Dockerfile}, left = 100pt of server3] (file) {\faFile};
        \draw[-latex, dashed,] (file) -- node[above] {Custom Impl.} ++(2.55,0) ;

        \draw[line width=1pt] (server1.east) to (2,2.5) to (router);
        \draw[line width=1pt] (server2) to (router);
        \draw[line width=1pt] (server3.east) to (2,0.5) to (router);
    
        \draw[dashed,line width=1pt,latex-] (scanner) -- node[right,midway,label=right:{Config}] {\faGear}++(0,-1) node[below] {ZMap, QScanner, \ldots } ;
    
        \draw[line width=1pt] (router) to (scanner);
    \end{tikzpicture}%
    \caption{Test environment Docker setup. Each server implementation is hosted within its own container and thus isolated. Public (\eg{} from the \quicInteropRunner{}) or self-built containers can be used.}%
    \label{fig:test_environment}
\end{figure}
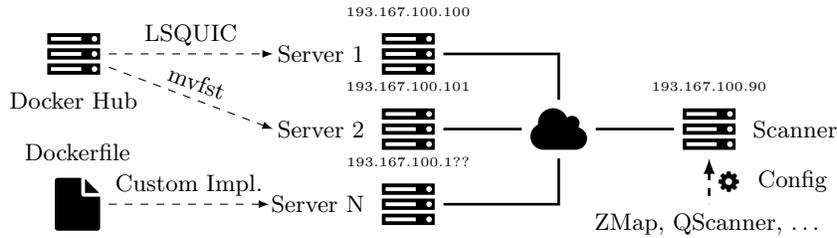

The environment is based on Docker and isolates different servers from the scanners (see \Cref{fig:test_environment}).
Each implementation is running in its own Docker container and is reachable via an individual IP address.
An additional container executes scanners, \ie{} \zmap{} and the \qscanner{}.
While individual containers can be added, we include all 16 existing server implementations of the \quicInteropRunner{}~\cite{seemann2020automating} as of October 2023 (see Appendix \Cref{tab:serverTestcaseSupport_withdomain} for a complete list).
The library developers themselves mainly provide and regularly update them.
Our test environment supports general \quic{} handshakes and HTTP/3 requests.

We used this environment \first{} to  test different scanners and configurations to improve \quic{} scans (\Cref{sec:scanning}) and \second{} to develop an approach to identify \quic{} libraries (\Cref{sec:identification}).
Furthermore, we publish this environment~\cite{toolchain} to allow others to reproduce our findings and update our identification approach in the future in case of library changes.
Updated \quic{} servers can easily be integrated using new and up-to-date containers published by the library maintainers or built individually.
Other researchers can use it to test scanners or scan configurations.

Besides our local environment, we collected a set of servers explicitly announced as test servers or operated by the developers of specific libraries.
They allow us to verify findings with real deployments on the Internet and  to analyze additional libraries where no \quic{} server is available for the \quicInteropRunner{} (\eg{} google.com for \gquiche{}~\cite{impl_QUICHE-Google}).
While only little information about a \quic{} library by Akamai is available, we confirmed with Akamai that they maintain their own library originally forked from \gquiche{}.

\section{Scanning for QUIC Deployments}%
\label{sec:scanning}
Scanning for \quic{} deployments is often a two-step process: \first{} identifying targets with a stateless scan and \second{} evaluating the capabilities of targets with a stateful scan completing a handshake.
We deployed this approach on large scale, improved the relevant components, and evaluated the state of \quic{} servers on the Internet.

\begin{figure}[b]
    \centering
    \footnotesize
    \begin{tikzpicture}[
    entity/.style={draw, minimum width=2cm, minimum height=0.6cm},
    innerentity/.style={draw, minimum width=1.95cm,minimum height={0.5cm}, align=center},
    nameentity/.style={draw, minimum width=3.7cm,minimum height={0.5cm}, align=center},
    node distance=0.15cm
    ]%

    \node[innerentity] (qscanner) {\qscanner{}};
    \node[innerentity, above right = 12pt and 0pt of qscanner] (h3) {HTTP/3};
    \node[innerentity, below right = 12pt and 0pt of qscanner] (invalid) {\acs{alpn} invalid};

    \node[innerentity, above left = 12pt and 0pt of qscanner] (nosni) {no \acs{sni}};
    \node[innerentity, below left = 12pt and 0pt of qscanner] (sni) {\acs{sni}};

    \node[innerentity, below left = 12pt and 0pt of nosni] (zmap) {\zmap{}};

    \node[nameentity, left = 10pt of zmap] (v6) {\underline{IPv6}: Hitlist \cite{gasser2018clusters,zirngibl2022clusters},\\ AAAA records};

    \node[nameentity, below = 10pt of v6] (dns) {\underline{DNS}: \acs{czds},\acs{cctld}, \ldots};

    \node[nameentity, above = 10pt of v6] (v4) {\underline{IPv4}: Full /0};
 
    \draw[-latex, thick] (v6) -- (zmap);
    \draw[-latex, thick] (v4) -- (zmap);
    \draw[-latex, thick] (zmap) -- (nosni);
    \draw[-latex, thick] (zmap) -- (sni);

    \draw[-latex, thick] (dns) -- (sni);
    \draw[-latex, thick] (dns) -- (v6);
    \draw[-latex, thick] (nosni) -- (qscanner);
    \draw[-latex, thick] (sni) -- (qscanner);

    \draw[-latex, thick] (qscanner) -- (h3);
    \draw[-latex, thick] (qscanner) -- (invalid);

    \end{tikzpicture}%
    \caption{Scan setup to identify \quic{} deployments and used libraries. The \acs{alpn} \texttt{invalid} scan is explained in \Cref{sec:identification}. Ethics are covered in \Cref{sec:conclusion}.}%
    \label{fig:scan_setup}
\end{figure}
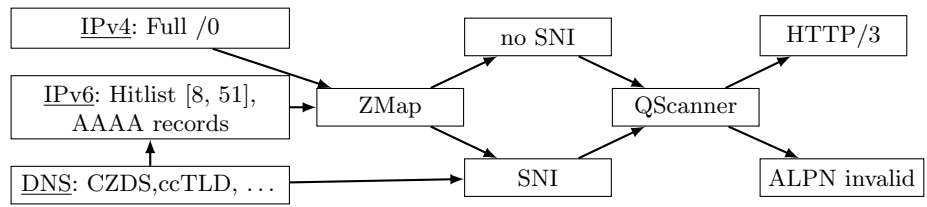

We conducted Internet-wide scans to analyze current \quic{} deployments, focusing on used libraries.
Our approach follows scan approaches based on \zmap{} to identify potential targets and \ac{dns} resolutions to find potential \ac{sni} values, followed by stateful \quic{} scans (see \Cref{fig:scan_setup}).

We seeded our \ac{dns} scans with more than \sm{400} domains, \eg{} from the \ac{czds}, a list of \ac{cctld} domains and domains extracted from \aclp{ctlog}.
We resolved all domains to A and AAAA records for IPv4 and IPv6 addresses.

For IPv4, we seeded \zmap{} with the entire address space.
Regarding IPv6, we used the IPv6 Hitlist~\cite{gasser2018clusters,zirngibl2022clusters} and added IPv6 addresses extracted from AAAA records. 
The latter ensures that we include IPv6 addresses of large \acp{cdn} in our dataset.
\citeauthor{zirngibl2022clusters}~\cite{zirngibl2022clusters} have shown that IPv6 prefixes of \acp{cdn}, \eg{} Fastly and Cloudflare, often appear aliased and are filtered by the regular hitlist service to prevent biases.
Therefore, additionally relying on AAAA records besides the Hitlist implements suggestions by the operators regarding proper usage~\cite{zirngibl2022clusters}.
We used the \qscanner{}~\cite{zirngibl2021over9000} for the stateful scan.
We conducted a \quic{} handshake to collect transport parameters and used
\textit{h3} as \ac{alpn} value.
After a successful handshake, we sent an \ac{http} request to collect \ac{http} header information.

In this work, we do not rely on \texttt{HTTPS} DNS resource records which could indicate \quic{} support via an \ac{http}/3 \ac{alpn}  value~\cite{rfc9460}.
\citeauthor{zirngibl2023svcb}~\cite{zirngibl2023svcb} have shown in March 2023 that the records are mostly used by Cloudflare so far.
Furthermore, we do not use Alt-Svc HTTP headers from TLS over TCP scans, an otherwise recommended approach to negotiate \ac{http}/3 support~\cite{web_almanac}.
On one side, they would require additional Internet-wide scans using TCP, on the other side, they are impacted by missing \ac{sni} values as well and not necessarily reveal additional targets.
Our approach solely relies on \quic{} probes sent out using \zmap{} and is independent of other scans. 

\subsection{ZMap: Version Negotiation}
The first important step is a \zmap{} scan, designed to effectively evaluate on a large scale whether a target IP address is offering a service on a given port~\cite{durumeric_zmap_2013}.
While \ac{tcp}-based protocols are identified based on the 3-way handshake (sending a SYN packet and expecting a SYN-ACK), the \ac{udp}-based \quic{} protocol requires a meaningful payload during the first flight.
Previous work~\cite{ruethQUIC2018, zirngibl2021over9000} relied on a payload that triggers a \ac{vn} by the server, \ie{} sending a \quic{} packet with an unsupported version.  
The \quic{} standard \cite{rfc9000} explicitly reserves versions following the pattern \texttt{0x?a?a?a?a} to be used to trigger a \ac{vn}.

In the following, we focus on the most recent module by \citeauthor{zirngibl2021over9000}~\cite{zirngibl2021over9000}.
It sets a reserved version and a valid size but omits most of the remaining information, \eg{} a valid Client Hello.
Servers should first parse the version field and directly send a \ac{vn} in case of an unsupported version~\cite{rfc9000}, allowing to identify \quic{} servers in a stateless manner. 
We tested this module in our test environment and found that three implementations do not respond to the probe, namely Amazon's \sTWOnquic{}, \lsquic{} and \aioquic{}.
They try to parse the entire initial packet first, only check the version afterward and do not respond to the existing module.
Therefore, deployments based on these libraries are not detected by the \zmap{} module and are missed by related work.

In contrast, sending a \quic{} initial packet including a valid Client Hello results in a response by all three implementations in our test environment.
However, it triggers a handshake and key exchange, thus creating state on the server.
To reduce state during \zmap{} scans, we extracted a valid initial packet from a \quic{} communication as raw bytes, but manually set a reserved version.
This extracted and customized packet can be used as pre-defined payload in combination with the existing UDP \zmap{} module.
It triggers a \ac{vn} which can be implemented without state on the server (cf.~\cite{rfc9000}) and works with all tested libraries.
We have published the required configuration and payload along with our test environment~\cite{toolchain}. 

We compared the three possibilities within an Internet-wide scan: \first{} a \ac{vn} with zero bytes, \second{} a \quic{} initial with version 1, and \third{} a \quic{} initial with a reserved version.
\Cref{tab:stats_zmap} shows results from a scan during October 2023.
The approaches with zero bytes or a proper initial packet result in similar sets of identified deployments.
However, our newly proposed approach with a valid \quic{} initial and a reserved version identified \sm{2.6} more IPv4 and \sm{7.4} IPv6 targets.
These targets are operated by few \acp{as}, mainly Amazon.
While \sTWOnquic{} responded to initial packets with a correct version in our test environment, Amazon deployments do not respond but probes time out.
This is due to missing \ac{sni} values as shown in the next section.

\textit{\textbf{Key take-away:}
Properly selecting  a \quic{} probe during \zmap{} scans can drastically influence the number of identified deployments.
Our test environment helps to identify high-quality probes and can be used in the future to adapt scans to potential library or protocol changes.
}

\begin{table}[t]
    \setlength{\tabcolsep}{1pt}

    \begin{minipage}{0.48\textwidth}
            \centering
            \small
            \caption{Found \quic{} deployments based on \zmap{} with three different probes in October 2023.}%
            \label{tab:stats_zmap}%
            \begin{tabular}{lrrrrrr}
                \toprule
                & \multicolumn{2}{c}{VN} & \multicolumn{2}{c}{Initial} & \multicolumn{2}{c}{Initial-VN} \\
                \cmidrule{2-3}\cmidrule{4-5}\cmidrule{6-7}
                IP & Addr. & ASes & Addr. & ASes & Addr. & ASes \\ 
                \midrule
                v4 & \sm{9.2} & \sk{7.5} & \sm{9.3} & \sk{8.7} & \sm{11.9} & \sk{8.7}  \\
                v6 & \sk{460.7} & \sk{2.7} & \sk{483.6} & \sk{2.9} & \sm{7.9} & \sk{2.9} \\
                \bottomrule
            \end{tabular}
    \end{minipage}%
    \hfill
    \begin{minipage}{0.48\textwidth}
            \centering
            \small
            \caption{Successful handshakes based on the \qscanner{}. For \ac{sni}, IP addresses with at least one successful domain are counted.}%
            \label{tab:stats_gocanner}%
            \begin{tabular}{lrrrrrr}
                \toprule
                & \multicolumn{2}{c}{No SNI} & \multicolumn{2}{c}{SNI} & \multicolumn{2}{c}{Total}\\
                \cmidrule{2-3}\cmidrule{4-5}\cmidrule{6-7}
                IP & Addr. & ASes & Addr. & ASes & Addr. & ASes \\ 
                \midrule
                v4 & \sk{522.6} & \sk{6.0} & \sk{411.9} & \sk{3.9} & \sk{876.6} & \sk{7.8} \\
                v6 & \sk{118.4} & \sk{2.4} & \sm{2.1} & \sk{1.3} & \sm{2.2}  & \sk{2.8}\\
                \bottomrule
            \end{tabular}
    \end{minipage}
\end{table}

\subsection{The Importance of SNI}%
\label{sec:sni}

Servers often host multiple domains on the same IP address.
The client can add the desired domain as \ac{sni} to the Client Hello during the \ac{tls} handshake.
Therefore, the server can use the correct certificate and authenticate itself.
In case no \ac{sni} is provided, the server can \first{} serve a default certificate, \second{} respond with an error, or \third{} time out the connection attempt.

Analyzing uniquely identified deployments by the third \zmap{} approach (see \Cref{tab:stats_zmap}) reveals that implementations require a proper \quic{} initial packet, respond with a version negotiation in case of an incorrect version, but do not respond to a \quic{} initial without \ac{sni}.
Amazon owns most of these addresses.
Thus, we searched for a domain hosted by them and conducted a handshake with \sk{1} IP addresses revealed by \zmap{} and the domain set as \ac{sni}.
In this scenario, the handshake is successful with \sperc{97.7} of the addresses.
Missing values result in timeouts and not necessarily \quic{} errors that can be analyzed.
This indicates the importance of \ac{sni} values during \quic{} handshakes.

To further investigate this, we scanned all identified \quic{} deployments with the \qscanner{} without \ac{sni} values and all addresses we could map to a domain based on \ac{dns} with \ac{sni}.
The latter set covers fewer IP addresses because we could not map domains to all addresses.
For IPv4, \zmap{} discovers \sm{11.8} addresses that react to our new probe.
Using the \qscanner{}, handshakes with \sk{522.6} addresses without \ac{sni} are successful (see \Cref{tab:stats_gocanner}). 
Considering the scan with \ac{sni} values \sk{601.9} addresses were tested.
At least one handshake with \sk{411.9} (\sperc{68.4}) addresses was successful.

As indicated by the total (\sk{836.2}), the overlap of successful addresses between \ac{sni} and no \ac{sni} scans is small.
Out of the \sk{522.6} IPv4 addresses with a successful handshake without \ac{sni}, we could map \sk{63.1} to a domain.
The handshake with an \ac{sni} was also successful for \sk{58.0} (\sperc{91.0}) out of these addresses.
In contrast, only \sperc{14.1} out of \sk{411.9} addresses with a successful \ac{sni} handshake were also successful without an \ac{sni} value.
Without \ac{sni}, they often resulted in generic \ac{tls} errors or timeouts.
This shows the importance of \ac{sni} for many deployments.

We used addresses from AAAA records besides the IPv6 Hitlist during our scans.
Therefore, we can identify more \quic{}-capable targets than the hitlist (\sm{7.9} compared to \sk{320}).
Similarly to IPv4, scans without \ac{sni} resulted in more timeouts.
For \ac{sni}, the increased number of IPv6 addresses compared to IPv4 was mostly due to Amazon which often responds with eight AAAA records compared to at most four A records per domain.  
The distribution of IPv6 addresses is more biased toward a few \acp{as}.
\sperc{73.7} of addresses were within Amazon (AS16509), the main contributor to fully responsive prefixes~\cite{zirngibl2022clusters}.

Comparing these results to previous findings by \citeauthor{zirngibl2021over9000}~\cite{zirngibl2021over9000} shows a significant increase in found \quic{} deployments (from \sm{2.1} to \sm{11.8}), their \ac{as} distribution (from \sk{4.7} to \sk{8.1}) and the number of successful handshakes (from  \sk{148.2} to \sk{503.1}).
The comparison is made based on IPv4 \zmap{} scans and handshakes without \ac{sni}, but the same effects can be seen for IPv6 and scans with \ac{sni}.
Two deployments mostly dominate the new addresses.
\first{} Akamai (AS16625) with \sm{5.6} addresses, a 20-fold increase, and \second{} Amazon (AS16509) with \sm{2.2} addresses not mentioned in 2021~\cite{zirngibl2021over9000}.
The deployment by Amazon is not necessarily a new deployment, but only detected now due to the different \zmap{} probe.
In comparison, the deployment by Akamai could also be detected with the previously existing \zmap{} \quic{} module and is thus new~\cite{akamai_deployment}.

\textit{\textbf{Key take-away:}
Scanning with or without \ac{sni} drastically influences results.
\ac{sni} values are essential with some operators, especially \acp{cdn} hosting multiple domains on the same IP address, to guarantee successful handshakes.
In some cases, \eg{} Amazon, incorrect or missing \ac{sni} values result in timeouts.
This leads to incorrect conclusions regarding the existence of \quic{} deployments.
}

\section{Library Identification}%
\label{sec:identification}

As shown in \Cref{sec:scanning} and by related work, different \quic{} libraries show different behavior even though they implement the same standards.
Based on these differences, we developed an approach to actively identify \quic{} libraries.

\subsection{Identification Methodology}

Our goal was to identify libraries with few packets and a high success rate.
Therefore, we used our test environment to evaluate different scanner configurations (\eg{} different \ac{tls} parameters) and the individual behavior of server implementations.
Our tests resulted in an approach relying on \quic{}-specific properties, namely \texttt{CONNECTION\_CLOSE} frames and transport parameters.
Appendix \Cref{tab:serverTestcaseSupport_withdomain} provides information which methodology identifies which library.

\begin{table}
    \setlength{\tabcolsep}{4pt}
    \centering
    \footnotesize
    \caption{Example error messages in response to the \acs{alpn} value \texttt{invalid}. The first column indicates who formats this exact error message. Full list:~\cite{toolchain}.}%
    \label{tab:scanMod_QScanner_ALPNOccurence}%
    \vspace{-0.5em}
    \begin{threeparttable}
        \begin{tabularx}{\textwidth}{cllX}
            \toprule
            & Implementation & Code & Message \\
            \midrule
            \parbox[t]{2mm}{\multirow{3}{*}{\rotatebox[origin=c]{90}{Scanner}}} & \quiche{}, \quicGo{} & \texttt{0x178} & tls: no application protocol \\
            \cmidrule{2-4}
            & \multirow{2}{*}{\xquic{}} & \multirow{2}{*}{\texttt{0x178}} & ALPN negotiation failed. Server didn't offer any protocols\\%
            \midrule
            \parbox[t]{2mm}{\multirow{7}{*}{\rotatebox[origin=c]{90}{Server library}}} & \multirow{2}{*}{\lsquic{}} & \texttt{0x178} & no suitable application protocol \\
            && \texttt{0x150} & TLS alert 80\tnote{1} \\
            &\nginx{} & \texttt{0x178} & handshake failed \\
            &\multirow{2}{*}{\quant{}} & \multirow{2}{*}{\texttt{0x178}} & (frame type: 0x6): PTLS error 120 (NO\_APPLICATION\_PROTOCOL) \\
            &\quinn{} & \texttt{0x178} &  peer doesn't support any known protocol \\
            &\haskellQuic{} & \texttt{0x178} & no supported application protocols \\
            \bottomrule
        \end{tabularx}
        \begin{tablenotes}
            \item[1] \lsquic{} checks \acs{sni} before \acs{alpn} and sends this error if \acs{sni} is required/incorrect.
        \end{tablenotes}
    \end{threeparttable}
\end{table}

\textbf{Error Message}
The \texttt{CONNECTION\_CLOSE} frame can provide a detailed error message, \ie{} a string without a specified structure.
We expect that individual deployments do not change this behavior, and the respective string is specific for a library.
We identified an unusual \ac{alpn} value as a reliable method to trigger an error during the handshake.
\ac{alpn} values are simple strings without limitations and thus allow us to send \textit{invalid}, a value we argue is highly unlikely used.

We found that nine implementations send a unique error message as response to unknown \ac{alpn} values.
\Cref{tab:scanMod_QScanner_ALPNOccurence} shows error messages of some implementations in our test environment and from public servers.
While all libraries indicate the same error, they send different text formats and slightly different content.
Seven implementations do not send a specific error message, and the \qscanner{} outputs a message from \quicGo{} internally.
Interestingly, \xquic{} continues with the handshake but does not include an ALPN extension, resulting in the \qscanner{} terminating the connection.
The \mvfst{} container image does not send an error but only indicates an error in internal logs.
However, testing against \texttt{facebook.com} results in a unique error message.
The same error is visible in the server logs in our local environment but most likely not sent due to a different configuration.
\lsquic{} checks the \ac{sni} value first and \ac{alpn} second.
However, both error messages are unique compared to others in our experience and identify \lsquic{}. 
We saw matching error messages for the remaining implementations with a server in our local environment and a test server on the Internet.

We verified the respective messages are created within the published code and found that messages have not been changed within the last three to five years for most libraries.
For example, the error message within \lsquic{} has not been updated for at least five years as of January 2024~\cite{lsquic_error}.
Thus, we expect them to remain stable in the future as well. 

\begin{table}
    \centering
    \setlength{\tabcolsep}{4pt}

    \footnotesize
    \caption{Examples of transport parameter~\cite{ianaQUICregistry} orders and \ac{tls} extensions~\cite{ianaTLSextensionregistry} to identify \quic{} libraries. The last two randomize their order. Full list:~\cite{toolchain}.}%
    \label{tab:transport_parameter}%
    \begin{tabular}{lll}
        \toprule
        Impl & Ext. & Transport Par.\\
        \midrule
        \multirow{2}{*}{\sTWOnquic{}} & 43-51 & 4-6-7-8-0-f \\
        & 51-43 & 4-6-7-8-0-f \\

        \lsquic{} & 51-43 & 4-6-7-8-0-f-2 \\
        
        \haproxy{} & 43-51 & 0-2-f-3-4-6-7-8 \\
        \mvfst{} & 43-51 & 0-6-7-4-8-a-3-2-f \\
        \bottomrule
    \end{tabular}
    \hfill
    \begin{tabular}{lll}
        \toprule
        Impl & Ext. & Transport Par.\\
        \midrule
        \quiche{} & 51-43 & 0-3-4-6-7-8-a-b-f \\

        \multirow{2}{*}{\nginx{}} & 51-43 & 4-8-6-7-3-b-a-0-f-2 \\
         & 43-51 & 4-8-6-7-3-b-a-0-f-2 \\

        Google Q. & 51-43 & set(0, 2, 3, 4, 6, 7, 8, f) \\
        Akamai Q. & 43-51 & set(0, 2, 3, 4, 6, 7, 8, f) \\
        \bottomrule
    \end{tabular}
\end{table}

\textbf{Transport Parameters}
Besides the error message usable for nine libraries, we use the order of transport parameters.
Our test environment reveals that most implementations send specific transport parameters in a particular order.
The explicit transport parameters are implemented and set by individual \quic{} libraries and are not necessarily part of the \ac{tls} library, \eg{} BoringSSL.
We create an identifier for each \quic{} library representing the parameters and their order.
We remove their values as they can be more easily configured.
In 2021, on- and off-net deployments of Facebook and Google sent the same set of parameters but with different values~\cite{zirngibl2021over9000}.
\Cref{tab:transport_parameter} shows examples for different libraries.

While the order of most implementations is stable, we found that four libraries randomize the order of a constant set (\eg{} \quant{}~\cite{quant_randomization} and \gquiche{}~\cite{google_randomization}).
However, the set alone can not be used as an identifier.
One permutation of \gquiche{} and \akaquic{} collides with \quinn{} and \haproxy{} respectively.
Nevertheless, the latter libraries do not randomize the order.
Therefore, a second handshake can be used to check whether permutations are visible or the identifier remains stable.
Furthermore, for some implementations, the set of transport parameters and their order is the same, \eg{} for \xquic{} and \haskellQuic{}.
Therefore, we add the order of the \textit{key share} and \textit{supported versions} extensions from the received Server Hello.
Both extensions must be present in \ac{tls} 1.3, and thus do not depend on a deployment-specific configuration.

\textbf{Additional Considerations}
During the development of our approach, we tested the following features but excluded them:
Additional \ac{tls} parameters could be used.
However, \quic{} relies solely on \ac{tls} 1.3, thus a simplified feature set compared to TCP/TLS stacks.
No specific features from earlier TLS versions and only four instead of hundreds of ciphers can be used~\cite{rfc9001,iana_ciphers}.
Furthermore, features are often unique to the \ac{tls} library.
Some libraries are shared by implementations and information is not necessarily unique.
Furthermore, information can be hardware (\eg{} preferred cipher suites) or configuration dependent (\eg{} supported groups).
While this can be used as information to differentiate specific deployments, it can not be used to identify libraries.
Different \quic{} features are not visible during the initial handshake (\eg{} the acknowledgement frequency or a path MTU discovery) or not necessarily static.
For example, we find 48 different configurations of transport parameter values across all \quicGo{} deployments but all with the same order.
We leave an extension to our approach to future work.

\textit{\textbf{Key take-away:}
    \quic{} libraries send different error messages or transport parameter orders and can therefore be identified.
    While these identifiers can change in the future, our published test environment can be used to identify new error messages testing against updated server implementations.
}

\subsection{Library Classification on the Internet}
We updated our scans (see \Cref{sec:scanning}) and used the \qscanner{} with two configurations: \first{} sending a valid \texttt{h3} \ac{alpn} and \second{} sending \texttt{invalid}.
We focused on the classification of libraries per IP address and expect a consistent \quic{} deployment on a reachable target.

\textbf{Stability of Identified Libraries}
We verified this assumption based on the stability of our identification.
Some targets can have multiple identifications due to multiple domain names mapping to a single IP address.
Within a single scan, we saw no inconsistent libraries per IP address, even though up to \num{100} domains were used as \ac{sni} values for each IP address.
Furthermore, we evaluate the stability of our identification over nine months comparing our library identification of four scans, one from April, July, October and December 2023.
\sm{6.6} IPv4 targets were visible in all four scans and were identified in at least one.
\sm{6.3} were identified in all four scans; for \sperc{99.99} of them, the identified library was consistent.
It shows that the identification is consistent and stable and not influenced by network or timing events, \eg{} caching.
Only for 215 IPv4 addresses changes were visible.
They are distributed across 53 different \acp{as} and consist of different library combinations.
We argue that our identification is stable and not influenced by network or timing effects. 
Changes are due to actual deployment changes.
For the remaining evaluation, we focused on the scan from October 2023.

\textbf{Resolving Collisions}
For 462 IPv4 addresses, the order of transport parameters either matches \haproxy{}/\quinn{} or the respective permutation of \akaquic{}/\gquiche{}, which randomize the order (see \Cref{tab:transport_parameter}).
However, for \num{146} (\sperc{31.6}) targets, the \textit{Error Message}-based approach identified the library as \quinn{} and directly resolved the collision (see \Cref{tab:scanMod_QScanner_ALPNOccurence}).
For the remaining \num{316} targets indicating a collision between \haproxy{} and \akaquic{}, we tried to conduct three additional handshakes and succeeded for \num{172}.
The remaining targets were not responsive anymore because the scans were conducted one week later.
Given the randomization, it is highly unlikely that multiple consecutive handshakes result in the same order.
Thus, for \num{171} targets, the order remained consistent, and we identified the library as \haproxy{}.
One target randomized the order during all three scans and was classified as \akaquic{}.

\begin{figure}[t]
    \includegraphics[trim={0 0.3cm 0 0},clip]{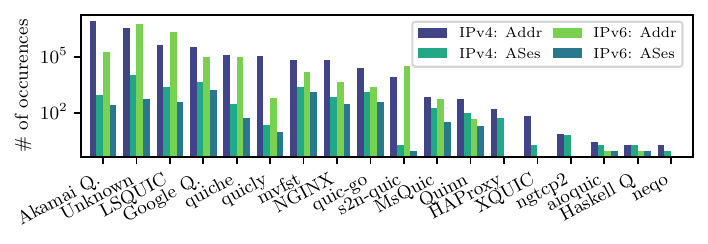}%
    \caption{Libraries on the Internet based on both scans (with/without \ac{sni}) and our approach. Handshakes are not necessarily successful with all targets.}%
    \label{fig:identification_results}
\end{figure}

\textbf{Approach Effectiveness}
In the IPv4 scan without \ac{sni}, we could identify the library for \sm{7.9} targets based on the \textit{Error Messages} and \sk{519.7} based on the transport parameters.
This shows the utility of the first approach.
We could identify a majority of target libraries because no successful handshake is required.
As shown in \Cref{sec:scanning}, many targets hosted by \acp{cdn} require a valid \ac{sni}, \eg{} Akamai. 
However, they terminate with a usable \textit{Error Message} in case of an invalid \ac{alpn} value.
While the \textit{Error Messages} approach identifies libraries on more targets, the transport parameter approach allows the identification of more distinct libraries.
It identified the library for \sk{167.2} previously unknown targets.
For \sk{352.5} targets, both approaches resulted in a matching classification, and in total \sm{8.0} (\sperc{67.1}) targets could be classified.  
Using a domain as \ac{sni} allowed identifying libraries for \sk{433.4} targets.
In this scenario, fewer results are based on the \textit{Error Messages} (\sk{289.8}) than the transport parameters (\sk{411.4}).
Combining both the results from the scan with and without \ac{sni} slightly improved the results, allowing us to identify the library on \sm{8.2} targets, \sperc{68.0} of initially identified deployments from \zmap{}.

The unknown category consists of \sm{3.9} IPv4 addresses.
However, \sm{2.2} (\sperc{56.7}) are hosted by Amazon and \sk{658.2} (\sperc{17.2}) by Cloudflare.
For both, scans without the correct \ac{sni} resulted in a timeout or a generic TLS error and due to load balancing, we can only map domains to few IP addresses.
We tested \sk{1} random IP addresses from Amazon with a valid \ac{sni} again (see \Cref{sec:scanning}), can conduct handshakes with \num{977} targets and identified the library as \sTWOnquic{}.
Therefore, we could identify more libraries, but opted to not use the same \ac{sni} with multiple millions of targets within the same provider for ethical reasons.
The remaining targets with unknown libraries mostly run into timeouts thus no information is available (another \sperc{5.8}), result in generic TLS errors or can not provide a certificate (\eg{} for scans without \ac{sni}).
We argue that for most of these targets, a valid \ac{sni} and thus a successful handshake results in required information to identify the library based on our approach.

\begin{figure}[t]
    \centering
    \footnotesize
    
    \begin{minipage}{0.54\textwidth}
        \includegraphics[trim={0 0.3cm 0 0},clip]{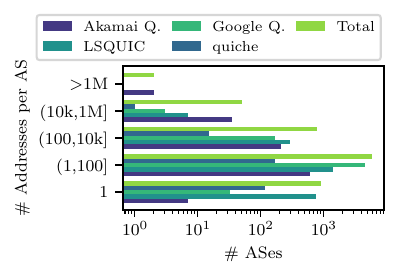}%
        \caption{Distribution of identified libraries in IPv4 targets across \acp{as}. Note the log x-axis.}%
        \label{fig:lib_distribution}
    \end{minipage}
    \hfill
    \begin{minipage}{0.44\textwidth}
        \includegraphics[clip]{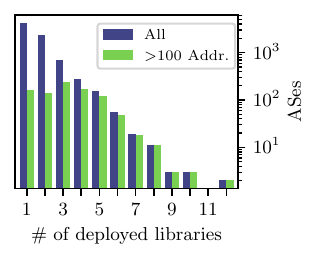}%
        \caption{Number of distinct libraries within ASes. Note the log y-axis.}%
        \label{fig:libs_per_as}
    \end{minipage}
\end{figure}

\textbf{Deployed Libraries}
\Cref{fig:identification_results} shows the classification for the IPv4 and IPv6 scans, including the number of visible \acp{as}.
We found all libraries at least once, but \kwik{}. 
\Cref{fig:lib_distribution} shows the distribution of libraries across \acp{as}.
Most \acp{as} contain between 1 and 100 \quic{} deployments we can identify, but \acp{as} with more than 100 or even \sk{10} deployments are visible.
The most visible implementation for IPv4 is the library from Akamai, which is also the most present operator based on IP addresses deploying \quic{} in general.
It is the only library with more than \sm{1} deployments in two \acp{as} each (both operated by Akamai itself).
However, it is only visible in \num{864} \acp{as}, compared to the libraries of other large so-called hyper giants~\cite{labovitz2010hypergiants}, \eg{} Google (\sk{4.7} \acp{as}) and Facebook (\sk{2.5} \acp{as}).
Interestingly, \quicGo{} (\sk{1.3} \acp{as}) and \lsquic{} (\sk{2.5} \acp{as}) show the largest distribution across different networks besides the libraries of hyper giants.
They are used for production-ready \quic{} servers in Caddy and LiteSpeed.
Therefore, deployments based on these products can easily use \quic{}, \eg{}~\cite{hostinger_lsquic_blogPost, a2hosting_lsquic}.

\Cref{fig:libs_per_as} shows up to 12 libraries within a single \ac{as}.
In most \acp{as} only one (\sk{4.2}, \sperc{54.3}) or two (\sk{2.3}, \sperc{30.1}) implementations are visible.
Within Amazon (AS16509) (\sperc{66.9} \sTWOnquic{}, \sperc{23.7} \lsquic{}) and Digital Ocean (AS14061) (\sperc{93.3} \lsquic{}, \sperc{3.9} \quicGo{}) the highest variety of implementations is visible.
Thus, identifying an \ac{as} operator does not necessarily reveal the used library.
Amazon develops its own library, but visible servers also rely on different \quic{} libraries.
Our methodology is required to correctly identify an implementation.  

\textbf{\ac{http} Server Header}
We compare our identification to \ac{http} Server headers.
The header should represent the \ac{http} implementation, but \quic{} libraries often directly include \ac{http}/3. 
However, the value is only available for \sk{699.2} IPv4 addresses, \sperc{79.8} of targets with a successful \quic{} handshake and \sperc{8.5} of targets we can identify a library of.
Thus, it is no replacement for our approach, and we only used it to cross-check our identification.
Our scans show that if available, the Server header value often matches or relates to the identified library.
Thus, for most targets relying on \akaquic{}, the Server header is \textit{AkamaiGHost}, while \textit{gvs} is used with \gquiche{}.
Appendix \Cref{tab:server_header} shows two prominent header values for the top 10 libraries.
We mainly focused on deployments outside the \ac{as} of the respective developers and found the same results supporting the correctness of our approach.
However, the value does not always reveal the library and is only available after a successful handshake, 

\hspace{0.5em}
\textit{\textbf{Key take-away:}
    Our approach is stable and works for the majority of targets.
    Compared to a more incomplete approach based on \ac{http} Server headers, our error message-based approach succeeds because no successful handshake is required.
    We find deployments for 18 different \quic{} libraries on the Internet and 12 different libraries within a single \ac{as}, showing the \quic{} ecosystem diversity.
}
\section{Conclusion}%
\label{sec:conclusion}

We analyzed \quic{} deployments and used server libraries on the Internet.
We evaluated \quic{} scanning approaches and are able to detect previously unseen deployments based on a new ZMap scan approach.
Furthermore, we developed an effective approach based on \textit{Error Messages} and the order of transport parameters to identify \quic{} libraries.
Eighteen different libraries are in use with at least one target on the Internet, and up to 12 libraries are visible within a single \ac{as}.
We can identify the library used by more than \sm{8.0} IPv4 and \sm{2.5} IPv6 addresses.
\textit{Network analysts and researchers need to be aware of this and consider potential differences based on different \quic{} stacks.}

\textit{Impact on Research} The variety of seen \quic{} implementations shows that key goals of the standardization have been met, and libraries can be quickly developed and deployed.
However, this diversity potentially increases the complexity of the network and might influence research regarding performance and security.
Our library identification could extend recent studies, \eg{} to evaluate the impact of libraries on performance~\cite{bauer2023performancebenefits}, to analyze the spin bit~\cite{kunze2023spin} or to evaluate ECN in \quic{}~\cite{sander2023ecn} instead of relying on more incomplete HTTP server headers.

\textit{Test Environment}
We presented and published~\cite{toolchain} an environment for testing scanners against different \quic{} libraries.
During the development of our scan and library identification methodology, we mainly limited ourselves to this environment to reduce the impact on the network and deployed \quic{} servers.
The environment helps to easily test scanners against various implementations without impacting real-world deployments.
It can easily be used in the future to adapt our approach to potential library changes.

\textit{Malicious Use}
Our identification approach can expose vulnerable deployments, thereby offering more effective means to identify exploitable systems.
We have contacted developers of \quic{} libraries about our findings and the potential to identify their specific implementations.

\textit{Ethics}
During our Internet-wide scans, we follow community standards~\cite{PA16,menloreport,durumeric_zmap_2013}.
We scan at a limited rate, use a local blocklist based on opt-out requests, and clearly identify our measurement infrastructure in WHOIS, RDNS, and via a hosted website on the machines.
We have received no complaints.

\newpage
{
\textbf{Acknowledgements}
This work was partially funded by the German Federal Ministry of Education and Research under the project PRIMEnet, grant 16KIS1370.
}

\appendix

\begin{table}
    \centering
    \small	
    \caption{Functional servers available in our test environment (Env). GH indicates a Github repository. The listed domains can additionally be used as test servers for respective libraries on the Internet. We confirmed with Akamai that they maintain their own library forked from \gquiche{}. The last two columns indicate which methodology can be used to identify the respective library (see \Cref{sec:identification}). The transport parameter (TP) order allows identifying more libraries but requires a successful handshake.}%
    \label{tab:serverTestcaseSupport_withdomain}
    \begin{threeparttable}
    \begin{tabular}{l l c l c l}
        \toprule
        \multicolumn{2}{c}{Implementation} & Env &  Domain & ALPN & TP \\
        \midrule
        \quicGo{}       & GH: \gurl{quic-go/quic-go}            & \yes{} & \textit{interop.seemann.io}                      &        & \yes{}               \\
        \ngtcpTWO{}     & GH: \gurl{ngtcp2/ngtcp2}              & \yes{} & \textit{nghttp2.org}                             &        & \yes{}               \\
        \quant{}        & GH: \gurl{NTAP/quant}                 & \yes{} & \textit{quant.eggert.org}                        & \yes{} & \yes{}               \\
        \mvfst{}        & GH: \gurl{facebookincubator/mvfst}    & \yes{} & \textit{www.facebook.com}                        & \yes{} & \yes{}               \\
        \quiche{}       & GH: \gurl{cloudflare/quiche}          & \yes{} & \textit{cloudflare-quic.com}                     &        & \yes{}               \\
        \kwik{}         & GH: \gurl{ptrd/kwik}                  & \yes{} &                                                  & \yes{} & \yes{}               \\
        \picoquic{}     & GH: \gurl{private-octopus/picoquic}   & \yes{} & \textit{test.privateoctopus.com}                 &        & \yes{}               \\
        \aioquic{}      & GH: \gurl{aiortc/aioquic}             & \yes{} & \textit{quic.aiortc.org}                         & \yes{} & \yes{}               \\
        \neqo{}         & GH: \gurl{mozilla/neqo}               & \yes{} &                                                  &        & \yes{}               \\
        \nginx{}        & \rurl{quic.nginx.org/}                & \yes{} & \textit{quic.nginx.org}                          & \yes{} & \yes{}               \\
        \msquic{}       & GH: \gurl{microsoft/msquic}           & \probably{}\tnote{1}  & \textit{[*].sharepoint.com}       &        & \yes{}               \\
        \xquic{}        & GH: \gurl{alibaba/xquic}              & \yes{} &                                                  &        & \yes{}               \\
        \lsquic{}       & GH: \gurl{litespeedtech/lsquic}       & \yes{} & \textit{www.litespeedtech.com}                   & \yes{} & \yes{}               \\
        \haproxy{}      & GH: \gurl{haproxytech/quic-dev}       & \yes{} & \textit{www.haproxy.org}                         &        & \probably{}\tnote{2} \\
        \quinn{}        & GH: \gurl{quinn-rs/quinn}             & \yes{} &                                                  &        & \probably{}\tnote{3} \\
        \sTWOnquic{}    & GH: \gurl{aws/s2n-quic}               & \yes{} & \textit{[*].cloudfront.net}                      & \yes{} & \yes{}               \\
        Haskell Q.      & GH: \gurl{kazu-yamamoto/quic}         &        & \textit{mew.org}                                 &        & \yes{}               \\
        Google Q.       & GH: \gurl{google/quiche}              &        & \textit{google.com}                              & \yes{} & \probably{}\tnote{3} \\
        Akamai Q.       &                                       &        & \textit{akaquic.com}                             & \yes{} & \probably{}\tnote{2} \\
        \bottomrule
    \end{tabular}
    \begin{tablenotes}
        \item[1] The container only supports \texttt{hq-interop} handshakes but no \texttt{h3}.
        \item[2] The Transport parameter order of \haproxy{} can collide with one permutation of \akaquic{} (see \Cref{sec:identification}).
        \item[3]  The Transport parameter order of \quinn{} can collide with one permutation of \gquiche{} (see \Cref{sec:identification}). 
    \end{tablenotes}
\end{threeparttable}
\end{table}

\section{Full List of Test Servers}
\Cref{tab:serverTestcaseSupport_withdomain} lists all \quic{} servers integrated in our local test environment introduced in \Cref{sec:environment}.
We use already available Docker images for each implementation provided to the \quicInteropRunner{}~\cite{seemann2020automating}.
Furthermore, it lists \quic{} targets on the Internet used besides our local environment to extend and verify our insights.
These targets are either listed as test servers for implementations by the \ac{ietf} working group~\cite{quicImplementations2021} or official company pages by the developers of libraries.
The table further shows which methodology can be used to identify each library.
While the identification based on \ac{alpn} values can only be done for nine libraries, it does not require a successful handshake.
This is especially helpful for deployments that require an \ac{sni} value as shown in \Cref{sec:scanning} and \Cref{sec:identification}.

\section{Overview about seen HTTP Server Header Values}
\Cref{tab:server_header} shows the two most common \ac{http} Server header values for each implementation and IPv4 target.
The header can only be collected for targets with successful \quic{} handshakes and a successive \ac{http} request.
The analysis combines results from the scan without and with \ac{sni}.
For the latter, different server headers can be received and are combined as comma separated list.

\begin{table}
    \setlength{\tabcolsep}{4pt}
    \centering
    \small
    \caption{Most common HTTP Server header value for top 10 implementations. Note that the header is only available after successful handshakes and \ac{http} requests. An asterisk indicates aggregated values containing the leading substring.}%
    \label{tab:server_header}
    \begin{tabular}{lrlS[table-format=6.0]r}
        \toprule
        & &     \multicolumn{3}{c}{HTTP Server Header} \\
        \cmidrule{3-5}
        Impl. & {Target} & Value & {Targets} & \% \\
        \midrule
        \multirow{2}{*}{\gquiche{}} &  \multirow{2}{*}{\num{213034}}& gvs 1.0 & 109571 & 51.43 \\
        &&gws & 39821 & 18.69 \\
        \midrule
        
        \multirow{2}{*}{\lsquic{}} &  \multirow{2}{*}{\num{194953}}& LiteSpeed & 193411 & 99.21 \\
        &&MCP\_VCLOUD\_LIVE & 283 & 0.15 \\
        \midrule
        
        \multirow{2}{*}{\quiche{}} &  \multirow{2}{*}{\num{121756}}& cloudflare & 104973 & 86.22 \\
        &&nginx & 9126 & 7.50 \\
        \midrule
        
        \multirow{2}{*}{\quicly{}} &  \multirow{2}{*}{\num{91602}}& Varnish & 88923 & 97.08 \\
        &&Cloudinary & 822 & 0.90 \\
        \midrule
        
        \multirow{2}{*}{\nginx{}} &  \multirow{2}{*}{\num{48875}}& nginx & 28160 & 57.62 \\
        &&GreyWS & 4906 & 10.04 \\
        \midrule
        
        \multirow{2}{*}{\quicGo{}} &  \multirow{2}{*}{\num{12526}}& Caddy & 11325 & 90.41 \\
        &&nginx & 426 & 3.40 \\
        \midrule
        
        \multirow{2}{*}{\sTWOnquic{}} &  \multirow{2}{*}{\num{7509}}& AmazonS3 & 1640 & 21.84 \\
        &&CloudFront & 1071 & 14.26 \\
        \midrule
        
        \multirow{2}{*}{\mvfst{}} &  \multirow{2}{*}{\num{5317}}& proxygen[*] & 5316 & 99.98 \\
        &&gunicorn, proxygen[*] & 1 & 0.02 \\
        \midrule
        
        \multirow{2}{*}{\akaquic{}} &  \multirow{2}{*}{\num{2376}}& Akamai[*]  & 1653 & 69.57 \\
        &&TLB & 87 & 3.66 \\
        \midrule
        
        \multirow{2}{*}{\msquic{}} &  \multirow{2}{*}{\num{927}}& Microsoft[*] & 805 & 86.84 \\
        &&Kestrel, Microsoft[*] & 52 & 5.61 \\
      
        \bottomrule
        \end{tabular}
\end{table}

\bibliographystyle{splncs04}
{\footnotesize\bibliography{reference,implementations}}

\end{document}